\documentclass[twocolumn,showpacs,preprintnumbers,amsmath,amssymb]{revtex4}
\usepackage{tabularx,graphicx}

\usepackage{color}
\usepackage{hyperref}
\hypersetup{
    colorlinks=true,
    linkcolor=blue,
    filecolor=blue,      
    urlcolor=blue,
}

\usepackage{color}

\newcommand{\bluee}{\textcolor{black}}

\usepackage{ulem}   

\begin{document}

\newcommand{\beq}{\begin{equation}}
\newcommand{\eeq}{\end{equation}}
\newcommand{\beqn}{\begin{eqnarray}}
\newcommand{\eeqn}{\end{eqnarray}}
\newcommand{\bmath}{\begin{subequations}}
\newcommand{\emath}{\end{subequations}}
\newcommand{\bra}[1]{\langle #1|}
\newcommand{\ket}[1]{|#1\rangle}

\title{Evidence against superconductivity in flux trapping experiments on  hydrides under high pressure $\&$ On magnetic field screening and expulsion in hydride superconductors}

\author{J. E. Hirsch$^{a}$  and F. Marsiglio$^{b}$ }
\address{$^{a}$Department of Physics, University of California, San Diego,
La Jolla, CA 92093-0319\\
$^{b}$Department of Physics, University of Alberta, Edmonton,
Alberta, Canada T6G 2E1}

\begin{abstract} 
It has recently been reported that hydrogen-rich materials under high pressure trap magnetic flux, a tell-tale signature of superconductivity \cite{e22}.
Here we point out that under the protocol used in these experiments the measured results indicate that the materials $don't$ trap
magnetic flux. Instead, the measured results are either experimental artifacts or originate in magnetic properties of the sample or its environment unrelated to superconductivity, Together with other experimental evidence analyzed earlier, this clearly indicates that these materials are not superconductors.

\noindent
{\bf In a second part, we discuss magnetic field screening and expulsion.}
\end{abstract}
\maketitle

\section{introduction}
Following the paper ``Conventional superconductivity at 203 kelvin at high pressures in the sulfur hydride system'', published in 2015 \cite{sh3}, 
several other hydrogen rich materials under high pressure have been reported in recent years to be high-temperature superconductors
based on observed drops in resistance versus temperature \cite{eremetsp,eremetslah,zhaolah,hemleylah,hemleylah2,yttrium2,yttrium,yttriumdias,thorium,pr,layh10,roomt,ceh,snh,bah,cah,cah2}.
Many more such materials have   been determined to be conventional 
high-temperature superconductors based on theoretical evidence \cite{semenok,review1,review2,theory1,theory2,theory3}. However,  little magnetic evidence has so far been provided
in support of the claims of superconductivity \cite{sh3,e2021p,e2022,nrs,huang}, and 
what evidence does exist has been strongly called into question \cite{hm3,hm4,hm7,huangmine}.

 In particular, these materials show no trace of magnetic {\it flux expulsion}, i.e. the Meissner effect,
 when cooled in the presence of a magnetic field \cite{sh3,e2021p,e2022}. They also apparently
 are able to screen very large applied magnetic fields  \cite{nrs}. This has been
 interpreted as indicating that the materials are ``hard superconductors'' with very strong pinning
 centers that prevent both flux penetration and flux expulsion  \cite{e2021p,e2022,nrs}. 
 We have argued that if that is the case the materials should also trap large magnetic fields \cite{hm5},
 and that observation of flux trapping  would provide definitive
  evidence that the materials can sustain persistent currents, hence are indeed superconductors         \cite{hm5}.
  
Experiments aimed at detecting flux trapping were recently performed by Minkov et al. and the results
analyzed and reported in Ref. \cite{e22}. Ref. \cite{e22} interprets the measured data as
clearly indicating that the materials are superconductors. Instead, we analyze here the information 
presented in Ref. \cite{e22}  and  conclude that it proves the absence of superconductivity in these
materials.

  \section{experimental protocols}
  The flux trapping experiments on sulfur hydride ($H_3S$) \cite{e22}  were performed under zero-field-cooling (ZFC) conditions
  for 13 values of applied field  ranging from 0 to $6T$, and under field cooling (FC) conditions for
  one field value only, $4T$. The results for both protocols for field 4T were reported to agree \cite{e22}.
  In the ZFC protocol,   the sample
  was cooled to low temperatures in zero magnetic field, a magnetic field was then applied and gradually
  increased to reach value $H_M$, then after 1 hour the external field was gradually decreased to zero, then  the resulting magnetic moment was
  measured with a SQUID magnetometer. 
  
               \begin{figure} []
 \resizebox{8.5cm}{!}{\includegraphics[width=6cm]{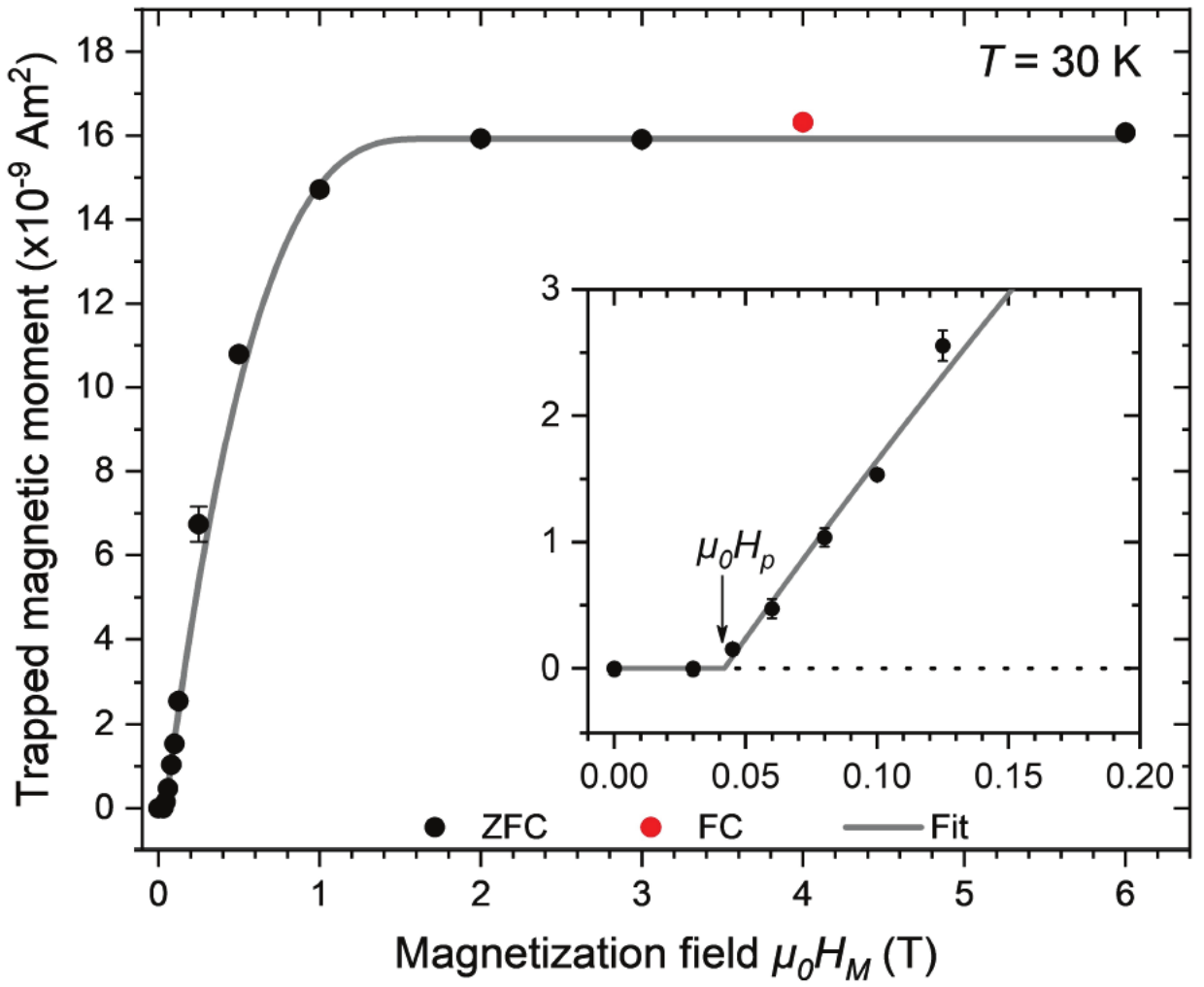}} 
 figt1-eps-converted-to
 \caption { Trapped magnetic moment for $H_3S$ at 30K, from Ref. \cite{e22}.
 The points are experimental data \cite{data}, the lines are a fit to the data performed in Ref. \cite{e22}.
}
 \label{figure1}
 \end{figure}

  Fig. 1 shows the experimental data and a theoretical fit to the data given in Ref. \cite{e22}.
  Note in particular that the measured magnetic moment rises linearly from zero when
  the applied field exceeds the threshold value $H_p$ both for the experimental data and
  for the theoretical fit.
  
    The experimental results were reportedly analyzed in Ref. 
   \cite{e22} assuming 
  the Bean model \cite{bean} controls the behavior of fields and currents in the material. 
From the experimental results, Ref. \cite{e22} inferred the parameters:

$H_p=0.042T$=threshold value of the applied field where it begins to penetrate the sample at low
  temperatures. Assuming
  demagnetization $1/(1-N)=8.5$, this implies a lower critical field value $H_{c1}=0.36T$.
  
    $H^*=0.835T$= minimum applied field that reaches the center of the sample 
    (called ``full penetration field''), with assumed sample 
    diameter and height
    $d=85 \mu m$, $h=2.5 \mu m$.
    
               \begin{figure} [t]
  \resizebox{8.5cm}{!}{\includegraphics[width=6cm]{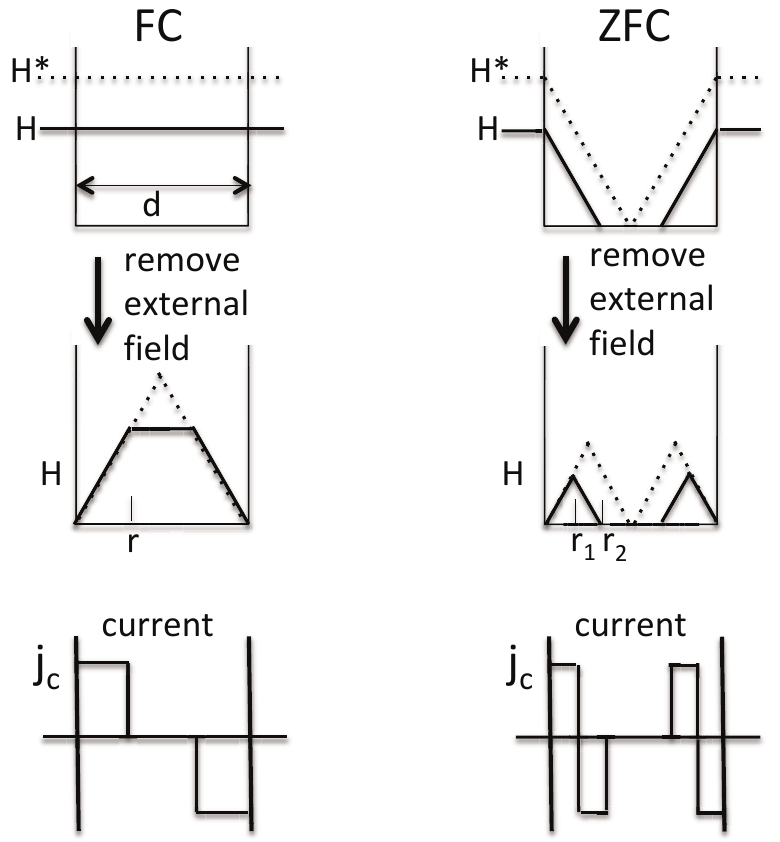}}
 \caption { Magnetic fields and currents predicted by the Bean model under field-cooled (FC) and
 zero-field-cooled (ZFC) protocols. Here we assume $H_p=0$ for simplicity.
}
 \label{figure1}
 \end{figure}

The measured moment
  was found to increase with magnetic field $H_M$ up to a maximum value
  of approximately $m_s=15.9\times 10^{-9}Am^2$ for $T=30K$ when the applied
  magnetic field was $\sim 1.7T\equiv H_M^{sat}$ or larger. Following the Bean model,
  Ref. \cite{e22} concluded that 
  \beq
  H_M^{sat}=2H^*+H_p
  \eeq
  from which the value of $H^*$ was extracted.
  
The theoretical fit performed in Ref. \cite{e22} assumed the magnetic moment is given by
(with $j_c$ the critical current)
\bmath
\beq
m=\int_{r}^{d/2} \pi r'^2 j_c h dr'=m_s[1-(\frac{r}{d/2})^3]
\eeq
\beq
r=r(H_M)=\frac{d}{2}(1-\frac{H_M-H_p}{2H^*})
\eeq
\emath
so that $r(H_p)=d/2$, $r(2H^*+H_p)=0$.
  
  \section{our analysis}
Just as Ref. \cite{e22}, we assume the validity of the Bean model. However we disagree that
Eqs. (2a), (2b) used by the authors of \cite{e22} is the proper way to calculate the trapped magnetic moment under ZFC conditions.
Instead, we argue that Eq. (2a) is the proper way to calculate the trapped  moment under FC conditions, provided Eq. (2b) is replaced by 
\beq
r=r(H_M)=\frac{d}{2}(1-\frac{H_M-H_p}{H^*})
\eeq
for $H_M<H^*+H_p$, $r=0$ for $H_M>H^*+H_p$, with $H_p=0$. This is illustrated in the left panels of Fig. 2. For ZFC conditions instead, the diagrams shown in the
right panels of Fig. 2 apply. For that case, the magnetic moment is given by
\beq
m=m_s[1-2(\frac{r_1}{d/2})^3+(\frac{r_2}{d/2})^3]
\eeq
where, for $H_M<H^*+H_p$
\bmath
\beq
r_1=\frac{d}{2}(1-\frac{H_M-H_p}{2H^*})
\eeq
\beq
r_2=\frac{d}{2}(1-\frac{H_M-H_p}{H^*}) .
\eeq
\emath
For $H^*+H_p<H_M<2H^*+H_p$, $r_1$ is given by Eq. (5a) and $r_2=0$, and for
$H_M>2H^*+H_p$, $r_1=r_2=0$.

           \begin{figure} [b]
  \resizebox{8.5cm}{!}{\includegraphics[width=6cm]{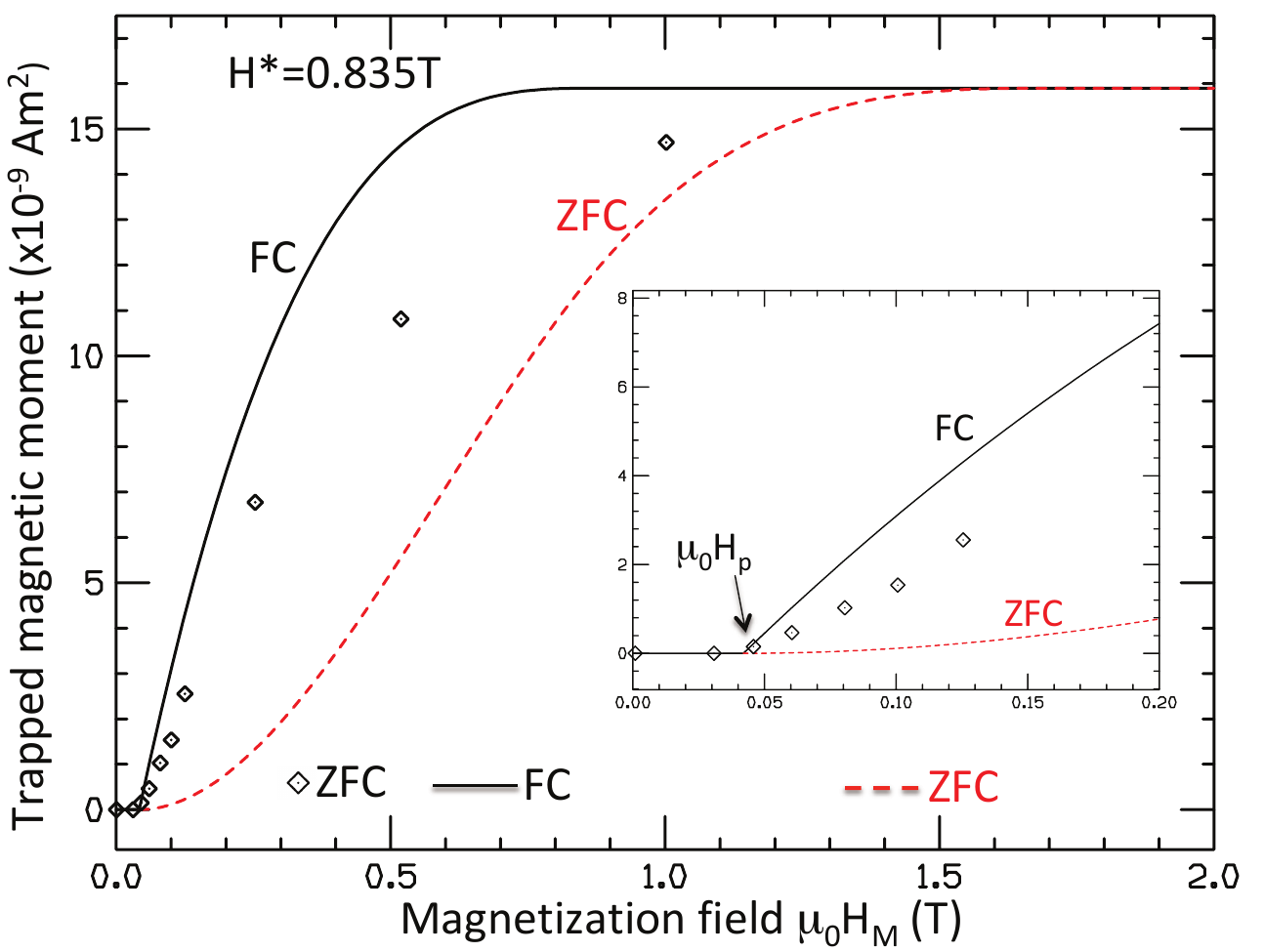}} 
 \caption {Expected trapped magnetic moment versus magnetization field $H_M$  for the parameters
 assumed in Ref. \cite{e22} under FC and ZFC protocols.
 For small $H_M$ the dependence is linear (quadratic) for FC (ZFC) protocols. The experimental points are also shown.
}
 \label{figure1}
 \end{figure}

Fig. 3 shows what these expressions predict for the trapped magnetic moment versus magnetization
field $H_M$ for the parameters
 assumed in Ref. \cite{e22}.
Most importantly, the moment rises from zero $linearly$ under FC conditions and
$quadratically$ for ZFC conditions. As seen in the inset, for
small fields the ZFC moment is very much smaller than the FC moment and  in stark  disagreement
with the experimental observations.

The experimental results of Ref. \cite{e22}  are actually well fit by our FC calculation for all values of the magnetization field $H_M$
if we take   the value of $H^*$ to be  twice as large as inferred in
Ref. \cite{e22}, i.e. $H^*=1.67 T$. This is shown in Fig. 4. We conclude that this agreement is accidental,
since the experimental protocol was ZFC for all but one experimental point \cite{e22}.

              \begin{figure} []
  \resizebox{8.5cm}{!}{\includegraphics[width=6cm]{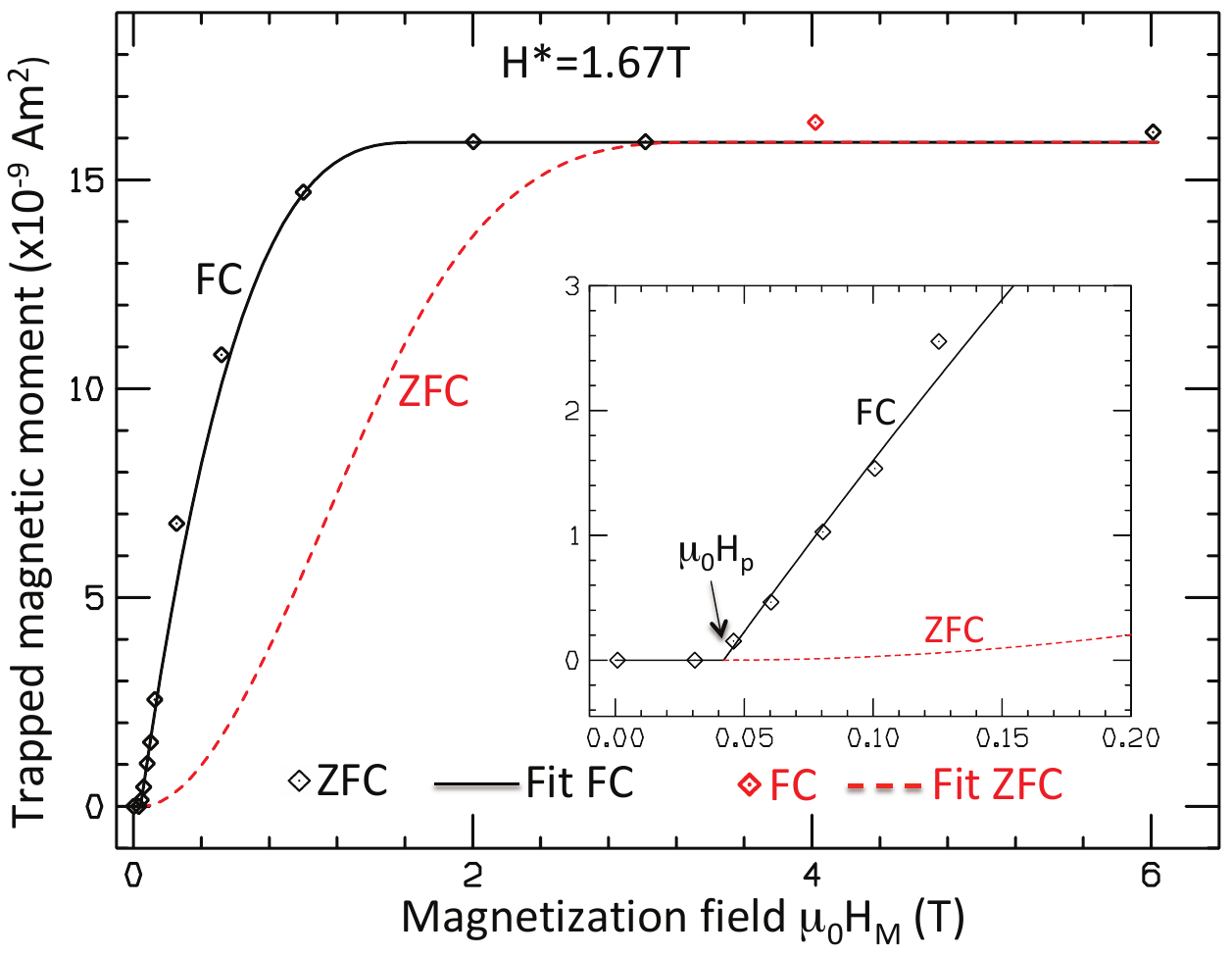}} 
 \caption {Expected trapped magnetic moment versus magnetization field $H_M$  
 assuming $H^*$ is twice the value inferred in   Ref. \cite{e22} ,
 i.e. $H^*=1.67 T$, compared to the experimental points.  
 Remarkably, the experimental points obtained with the ZFC protocol
 are actually fitted by the calculation assuming FC.}
 \label{figure1}
 \end{figure} 
 
 In order to try to fit the low field ZFC experimental data to the ZFC calculation we would have to take
a much smaller value of $H^*$. Fig. 5 shows the results for $H^*=0.2 T$, chosen to fit as well as possible
 the low field data. In addition to not fitting the low field data very well, the higher field data deviate strongly from the
theoretical ZFC curve. For this assumed value of $H^*$  the trapped moment saturates for $H_M^{sat}=0.44T$ (Eq. (1)),
in clear contradiction with the experimental data that show no saturation till $H_M>1T$.

                \begin{figure} []
  \resizebox{8.5cm}{!}{\includegraphics[width=6cm]{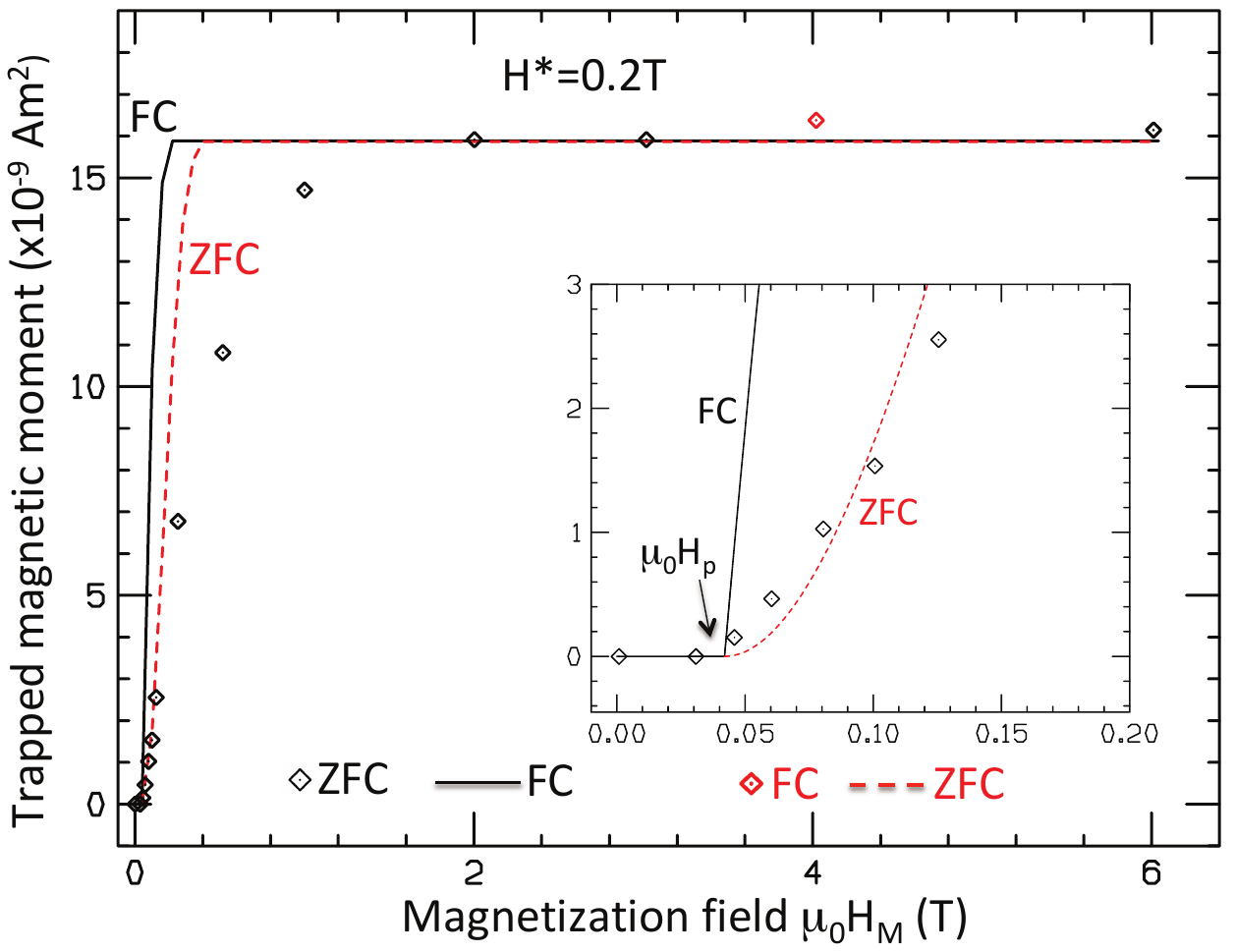}}
 \caption {Trapped magnetic moment versus magnetization field $H_M$  
 assuming $H^*=0.2T$ to fit the low-field experimental data. The ZFC curve saturates at $H_M^{sat}=0.44T$, well before the experimental data.}
 \label{figure1}
 \end{figure} 

\section{discussion}
Is it possible that under the ZFC protocol of the experiment  with the field $H_M$ applied for
1 hour, the field could penetrate sufficiently so as to mimic the FC protocol? It is not possible, because
Ref. \cite{e22} also measured the rate of flux creep and there was negligible flux creep
over a 1 hour period even at temperatures as high as 165K. Also, according to the 
NRS experiment \cite{nrs} the flux didn't penetrate over times substantially larger than 1 hour.

Therefore, the experimental results of Ref. \cite{e22} shown in Fig. 1 of this paper are incompatible
with the interpretation that the magnetic moment observed originates in flux trapping. If the
magnetic moment had originated in flux trapping, it would rise quadratically from zero as function of
the magnetization field $H_M$ under
the ZFC conditions of the experiment, not linearly as observed.
Therefore, the experiment indicates  that there is no  flux trapping in this material, $H_3S$.
As argued in Refs. \cite{hm7,hm5}, if the material doesn't trap flux, and in addition it does not
expel flux, then the material is not a superconductor.

The question then arises, what is the origin of the magnetic moments measured in 
Ref. \cite{e22} shown in Fig. 1? We suggest they are either experimental artifacts
associated with the experimental apparatus used (SQUID magnetometer) or 
magnetic moments of localized spins originating either in the sample or in the
diamond anvil cell environment (gasket, etc). It is also possible that the 
measurements could signal unexpected collective magnetic behavior of hydrogen-rich materials
under high pressure, as suggested in Ref. \cite{fm}.

 To confirm the results of our analysis we suggest that it would be of interest to repeat
 the measurements of Ref. \cite{e22} under FC conditions. We expect that the 
 results will be similar to the results under ZFC conditions, in contradiction with
 what is expected from trapped flux shown in Figs. 3 and 4, namely a marked difference between FC and ZFC behavior, and consistent with the hypothesis that
 the origin of the magnetic moments measured is localized spins rather than delocalized supercurrents.
 We suggest that it would also be informative to perform these experiments 
 using FC and ZFC protocols for a known hard superconductor and verify 
 the expected qualitatively different behavior shown in Figs. 3 and 4.

 Finally, we would   like to point out that the  interpretation of the measurements of magnetic moment 
 of Ref. \cite{e22}  as originating
 in flux trapping  with $H^*\sim 0.8T$ appear to be in contradiction with the magnetic moment measurements
 presented in Ref.  \cite{e2021p}. For example, according to the former (see our Fig. 2 top right panel) for an applied field
 $H\sim H^*/4=0.2T$ the magnetic field should still be excluded from more than $75 \%$ of the
 sample even  at temperature $T\sim 100K$ (see Fig. 1c of \cite{e22}). Instead, the magnetic moment measurements shown in Fig. 3a of  Ref. \cite{e2021p}
 indicate that the diamagnetism has essentially disappeared at that point.

  \begin{acknowledgments}
FM 
was supported in part by the Natural Sciences and Engineering
Research Council of Canada (NSERC) and by an MIF from the Province of Alberta.
We are grateful to the authors of Ref. \cite{e22} and particularly V. Minkov for
clarifying information.

\end{acknowledgments}

\noindent{\bf Second part starts on next page}

  \clearpage
    \begin{widetext}
    
        \section{second part}
        The above manuscript was posted on arxiv on 14 Jul 2022 and published in
        J Supercond Nov Magn 35, 3141–3145 (2022). What follows is a new manuscript, under consideration for publication in Nature Communications
        as  ``Matters Arising'',
        submitted to arxiv on 11/14/2022, that arxiv decided should be combined with the
        above manuscript to allow posting. \\
        \vskip0.2in
          \end{widetext}
            {\bf On magnetic field screening and expulsion in hydride superconductors}

              \begin{figure*} [t]
 \resizebox{16.5cm}{!}{\includegraphics[width=6cm]{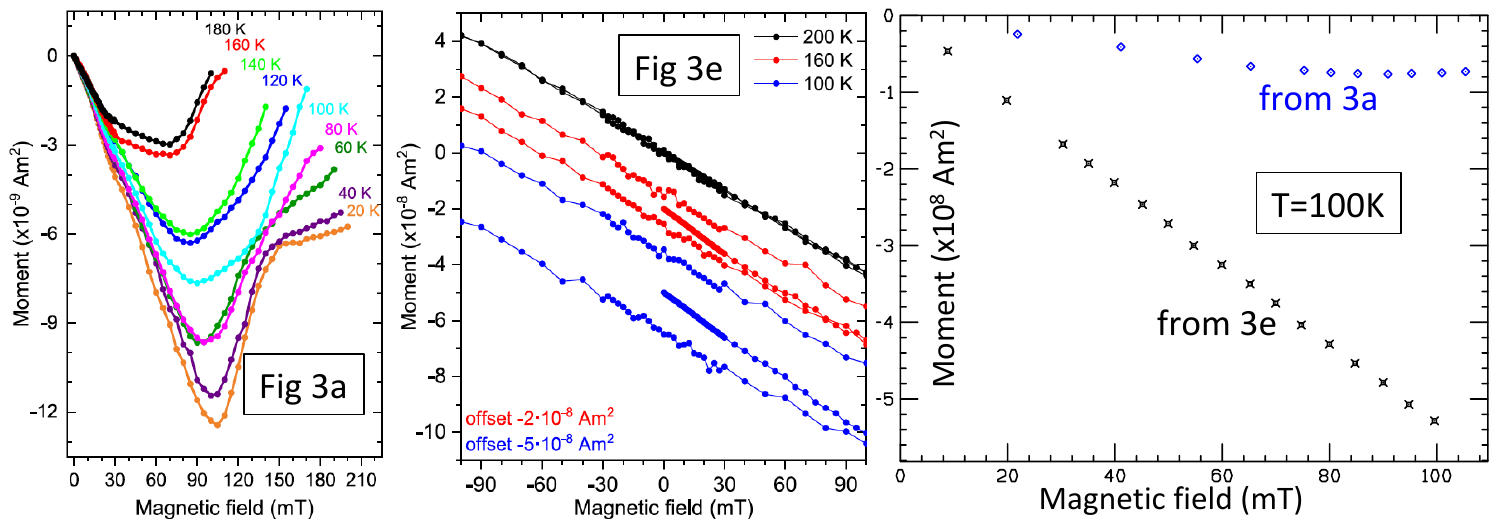}} 
 \caption {Left panel: magnetic moment versus applied field for $H_3S$ under pressure, from Fig. 3a of Ref.
 \cite{e2021pn}. Center panel: Magnetic moment versus applied field in a hysteresis cycle, from Fig. 3e of Ref.
 \cite{e2021pn}. The middle  blue curve in the center panel is presumably the virgin curve,
 which should be identical to the light blue curve on the left panel labeled 100K. Right panel: quantitative comparison of the
 virgin curves for 100K from the left panel (3a) and the center panel (3e). }
 \label{figure1}
 \end{figure*}

Ref. \cite{e2021pn} presents evidence for magnetic field screening and ``subtle'' evidence for magnetic field
expulsion in hydrides under high pressure, which is argued to support   the
claim that these materials are high temperature superconductors. We point out here that  data presented 
in  Ref. \cite{e2021pn} appear to be  inconsistent (i) with one another, \bluee{(ii) with other measurements reported by the same authors on
the same samples \cite{e2015n,e2021n}},  and (iii) with expected behavior of standard superconductors. 
This suggests that these magnetic phenomena reported for these materials are not associated with superconductivity, 
undermining the claim that these materials are high temperature superconductors.

In 2015, Eremets and coworkers reported high temperature superconductivity in sulfur hydride (hereafter $H_3S$) under pressure \cite{e2015n},
starting the hydride superconductivity epoch.
Since then to the present,  considerable evidence for superconductivity in various pressurized hydrides has been presented based on resistance measurements \cite{troyan22n}, however
little magnetic evidence for superconductivity  has been reported so far. In their original paper \cite{e2015n} Eremets and coworkers presented some
magnetic evidence   based on SQUID measurements. After a long hiatus, new evidence was presented 
this year in Nat. Comm. 13, 3194 (2022) \cite{e2021pn}. That evidence is the subject of this comment. 
We focus here on the magnetic measurements reported for sulfur hydride ($H_3S$),
but exactly the same considerations apply to the same measurements reported for 
lanthanum hydride ($LaH_{10}$) in Ref. \cite{e2021pn}, the only other hydride material for which magnetic measurements have been reported to date. 

Figure 6 left and center panels reproduce Fig. 3a and Fig. 3e of ref. \cite{e2021pn}. To the best of our understanding from carefully reading the
paper, both panels show in their light-blue and blue curves respectively the same quantity: magnetic moment versus magnetic field, for the same sample at the same
temperature (100K)  and same pressure (155 GPa). The middle blue curve in the center panel is the virgin curve, which starts (when properly shifted vertically,
as shown in Fig. S10 of \cite{e2021pn}) 
with zero moment for zero applied field. It should be the same as the light blue curve labeled 100K on the left panel. Yet the  curves look very different. The left panel curve shows an upturn for magnetic field
beyond 95mT while the center panel curve show no upturn. When plotting both curves on the same scale in the right panel
in Fig. 6 it is apparent that they  are very different in magnitude and shape. 


It should also be noted that the rapid decrease in the magnitude of the magnetic moments beyond the minimum points of the curves shown in  Fig.
6 left panel is inconsistent with what is expected for a type II superconductor with very large upper critical 
field \cite{tinkhamt2n}, estimated in Ref. \cite{e2021pn} to be $H_{c2}(T=0)\sim 97T$. For example, at $T=100K$ 
 $H_{c2}(T)$ should be above $60T$. When corrected for demagnetization factor estimated as
$1/(1-N)\sim 8.5$ in Ref. \cite{e2021pn}, it implies that the curve labeled $T=100K$ should evolve smoothly
from its value attained at $H \sim 95mT$ \bluee{approaching  zero at  or beyond} $H_{c2}(T)(1-N) \sim 7T$. 
This is qualitatively inconsistent with the
behavior seen in Fig. 6 left panel that shows that the   magnetic moment magnitude
has already decreased to less than $15\%$ of its maximum value for a field as small as $H\sim 0.2T\sim H_{c2}(T)/35$. 
\bluee{Furthermore, in the presence of strong pinning,  which Ref. \cite{e2021pn} claims has to exist in order to explain 
the absence of flux expulsion in their samples, the decay of the induced diamagnetic moment should be even slower than for
an ideal type II superconductor \cite{bean,senoussin}, hence very much slower than what is shown in Fig. 6 left panel.}


\bluee{We also point out that  the  magnitude of diamagnetic moment versus temperature under zero field cooling
reported  
in Ref. \cite{e2021pn} Figs. 2e and S1 left middle panel differs by a factor of 4 or more from the same quantity reported in 2015 in  Ref. \cite{e2015n} Figs. 4a and extended data Fig. 6c
 for samples estimated to be of similar size,
  with the earlier result showing the larger moment. While in field cooling experiments one may expect substantial
  variations in magnetic moment depending on sample quality, this is not expected to be the case for zero field cooling experiments.}

\bluee{Figure 7 shows as a blue curve the magnetic moment versus magnetic field at temperature 100K from the left panel
of Fig. 6, i.e. Fig. 3a of Ref. \cite{e2021pn}, compared with the magnetic moment versus magnetic field for a hysteresis cycle at the same temperature for the
same sample at the same pressure reported in 
Fig. 4a of Ref. \cite{e2021n}, that was used to obtain the
 critical current data shown in Fig. S5 of Ref. \cite{e2021pn}. The blue curve on the left panel of Fig. 7 should be the virgin curve for this hysteresis cycle,
joining smoothly the green curve, as is universally seen in such measurements for superconductors. One such typical example
is shown on the right panel of Fig. 7, from Ref. \cite{senoussin}. It can be seen that the blue curve on the left panel shows no hint of 
joining the green curve. In other words, these measured results on the same sample for the same temperature and pressure measured
in the same laboratory \cite{e2021pn,e2021n} are completely incompatible with one another under the assumption that they arise from
superconductivity in the sample.}

        \begin{figure} [t]
 \resizebox{8.5cm}{!}{\includegraphics[width=6cm]{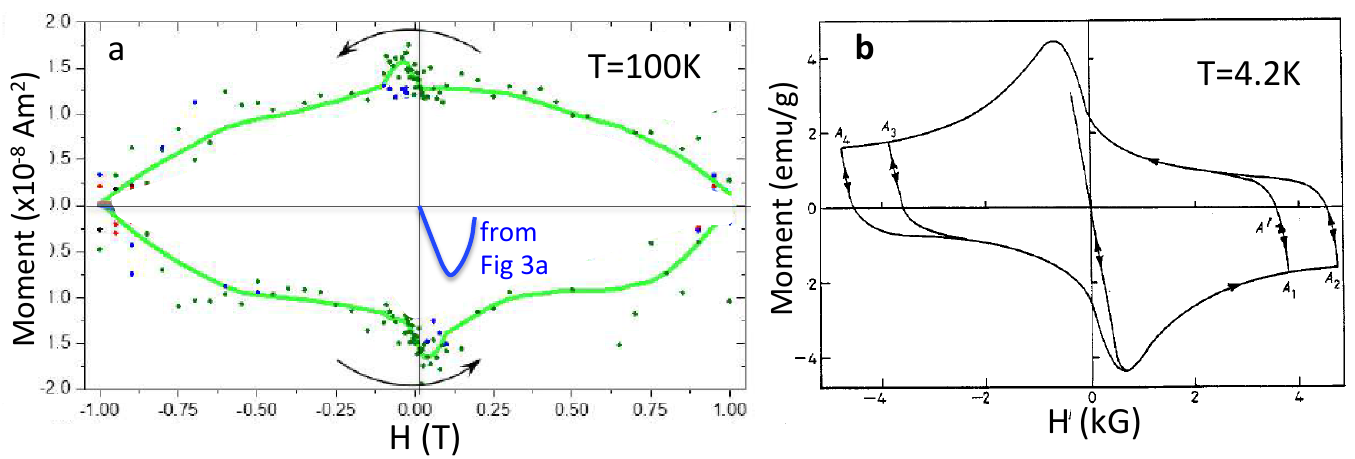}} 
 \caption {Green curve, left panel: hysteresis cycle for magnetic moment of $H_3S$ at 100K, from Fig. 4a of Ref. \cite{e2021n}. Those data were used
to obtain the
 critical current data shown in Fig. S5 of Ref. \cite{e2021pn}. The blue curve on the
 left panel shows the
 magnetic moment versus magnetic field for 100K from the light blue curve on the left panel of Fig. 6, which is Fig. 3a of Ref. \cite{e2021pn}.
 Right panel: a typical hysteresis cycle for a type II hard superconductor, from Ref. \cite{senoussin}. The virgin curve starting at the origin smoothly joins the
 hysteresis loop curve.}
 \label{figure1}
 \end{figure}

Ref. \cite{e2021pn} \bluee{says that it} uses a background subtraction procedure.
However, neither is the
background signal given in Ref. \cite{e2021pn} nor is the  procedure used  clearly explained. Perhaps, more information
on the data processing that has been performed would help explain some of the anomalies pointed out above. But even with such clarification we believe that the above analysis  indicates that the   reported magnetic measurements \cite{e2021pn} are inconsistent with the assumption that they originate 
in superconductivity. Instead, \bluee{we suggest that they} originate in localized magnetic moments associated with the samples, the diamond anvil cell
environment, and/or the measuring apparatus.

The signature property of superconductors, that cannot be mimicked by localized magnetic moments, is the Meissner effect, the ability  to $expel$ magnetic fields
when cooled in a field   (FC). In Ref. \cite{e2021pn},
the authors claim 
to find {\it ``subtle Meissner effect in FC measurements at 2 mT''} indicated by the 
 light blue  curve 
 \bluee{ shown in their Fig. S1 middle left panel. 
 However, when the same data are plotted in  Fig. SI1 middle left panel of Ref. \cite{e2021n} without the light blue curve, no 
 evidence for a Meissner effect is seen. }
While for some standard superconductors with strong pinning the percentage of flux expulsion 
 (Meissner fraction) can be very small
 for larger fields, it rapidly increases for small fields, as shown e.g. in Refs. \cite{meissner1n,meissner3n,thick2n,meissner4n}.
The Meissner fraction is expected to depend on the ratio $H/H_{c1}$ \cite{meissner2n},
 and  for $H_3S$ $H_{c1}$ is estimated to be $0.82T$ \cite{e2021pn}, which is more than an order of magnitude larger than
 lower critical fields for standard superconductors with high $T_c$ such as cuprates and pnictides.
 So the field 2mT of Fig. S1 of Ref. \cite{e2021pn} is equivalent to a field of less than  2 Oe for those other materials, for which a 
sizable Meissner fraction is found \cite{meissner1n,meissner3n,thick2n,meissner4n}. 
\bluee{It should also be noted that in Ref. \cite{e2015n} Extended Data Fig. 6 (c) the authors plotted magnetic moment under FC for magnetic
fields down to 0.2mT showing no evidence for a Meissner effect.}
Additionally, the Meissner fraction is expected to increase as the thickness of the sample decreases \cite{thick1n,thick2n}, 
and the samples used in these high pressure experiments are rather thin. 



\bluee{Elsewhere we have also called attention to the facts that (i) Fig. SI1 of Ref. \cite{e2021pn}  lower
left panel shows that the ZFC and FC magnetic moment curves for the precursor sample, not expected to be
superconducting, also show an unexplained divergence around 200K \cite{hmmren},  (ii)
the behavior of magnetic moment versus temperature  shown in Fig. 6 is incompatible with the 
claim of Ref. \cite{nrsn}, referenced in Ref. \cite{e2021pn} in support of superconductivity of sulfur hydride, that a magnetic field as large as 0.68T is excluded from  the sample \cite{hnrs}, and
(iii) ac magnetic susceptibility measurements  for sulfur hydride \cite{huangn} referenced in Ref. \cite{e2021pn} as evidence for 
superconductivity  
were shown to result from an experimental artifact \cite{huangminen}.}

\bluee{Recently, the authors of Ref. \cite{e2021pn} also reported measurement of trapped magnetic  flux in their samples
as evidence for superconductivity \cite{etrappedv2n}. We pointed out \cite{hmtrappedn} that the reported linear behavior of
trapped moment versus field in zero field cooling experiments \cite{etrappedv2n} is inconsistent with the expected behavior of
hard superconductors \cite{beann}. In addition, the magnetic moment measurements reported in Ref. \cite{e2021pn}
indicate that at low temperatures magnetic fields of up to 95mT are excluded from the sample (see Fig. 6 left panel here),
which is inconsistent with the reported finding in Ref. \cite{etrappedv2n} that applied fields as small as 50mT
penetrate and are trapped by the same samples.}

In summary, we argue that the matters pointed out here cast doubt on the interpretation of Ref. \cite{e2021pn} that the reported measurements
 originate in   superconductivity.



\begin{acknowledgments}
We acknowledge \bluee{some} helpful  correspondence with the authors of Ref.  \cite{e2021pn}.  
JEH is grateful to R. Prozorov for illuminating discussions. 
FM 
was supported in part by the Natural Sciences and Engineering
Research Council of Canada (NSERC) and by an MIF from the Province of Alberta.

\end{acknowledgments}
\medskip

\noindent {\bf Author contributions:} JEH initiated the study. JEH and FM analyzed the data and prepared
the manuscript.

\noindent {\bf Competing interests:} the authors declare no competing interests.

\noindent {\bf Data availability statement:} The data that support the findings of this study are available from the authors upon  reasonable request.

 \end{document}